%% file: main.tex
\documentclass[conference]{IEEEtran}
\IEEEoverridecommandlockouts
\usepackage{cite}
\usepackage{amsmath,amssymb,amsfonts}
\usepackage{algorithmic}
\usepackage{graphicx}
\usepackage{textcomp}
\usepackage{xcolor}

\usepackage[normalem]{ulem}

\def\BibTeX{{\rm B\kern-.05em{\sc i\kern-.025em b}\kern-.08em
    T\kern-.1667em\lower.7ex\hbox{E}\kern-.125emX}}
\begin{document}

\title{Measuring the Exploitation of\\ Weaknesses in the Wild
}

\author{\IEEEauthorblockN{Peter Mell}
\IEEEauthorblockA{\textit{Computer Security Division} \\
\textit{National Institute of Standards} \\
\textit{and Technology}\\
Gaithersburg, USA \\
0000-0003-2938-897X \\
peter.mell@nist.gov}
\and
\IEEEauthorblockN{Irena Bojanova}
\IEEEauthorblockA{\textit{Software and Systems Division} \\
\textit{National Institute of Standards} \\
\textit{and Technology}\\
Gaithersburg, USA \\
0000-0002-3198-7026 \\
irena.bojanova@nist.gov}
\and
\IEEEauthorblockN{Carlos Galhardo}
\IEEEauthorblockA{\textit{Dimel, Sinst} \\
\textit{Instituto Nacional de Metrologia}\\
Duque de Caxias, Brazil \\
0000-0002-7398-8182 \\
cegalhardo@inmetro.gov.br}
}

\maketitle

\begin{abstract}
\input{abstract}
\end{abstract}

\begin{IEEEkeywords}
attack, exploit, machine learning, weakness, vulnerability, security
\end{IEEEkeywords}

\input{body}


\input{references}

\end{document}

%% file: abstract.tex
Identifying the software weaknesses exploited by attacks supports efforts to reduce developer introduction of vulnerabilities and to guide security code review efforts. A weakness is a bug or fault type that can be exploited through an operation that results in a security-relevant error. 
Ideally, the security community would measure the prevalence of the software weaknesses used in actual exploitation. This work advances that goal by introducing a simple metric that utilizes public data feeds to determine the probability of a weakness being exploited in the wild for any 30-day window. The metric is evaluated on a set of 130 weaknesses that were commonly found in vulnerabilities between April 2021 and March 2024. Our analysis reveals that 92\,\% of the weaknesses are not being constantly exploited.

%% file: body.tex
\section{Introduction}
\label{Intro}

A software security weakness is a bug or fault type that can be exploited through an operation that results in a security-relevant error \cite{Bojanova2023}. 
The ``exploitation of a weakness'' refers to an adversary launching an attack that leverages a vulnerability with the underlying weakness. 

The development of metrics relative to the exploitation of weaknesses can identify the types of coding errors that lead to exploited vulnerabilities.
Such metric data will guide efforts to reduce the introduction of vulnerabilities by developers (e.g., with secure coding education).
It can also enhance both manual and automated code reviews by prioritizing the detection of the types of bugs being exploited.
A focus on \textit{exploitation} is crucially necessary because only ``2\,\% of published vulnerabilities have observed exploits in the wild'' \cite{CyentiaReport} and even fewer are being actively exploited.  

Ideally, the security community would measure the prevalence of software weaknesses used in exploitation in the wild (i.e., with actual attacks). 
Unfortunately, the data required for directly measuring these metrics are not publicly available. 

However, indirect and statistically-based measurements are attainable using data generated by both humans and machine learning. Complementary public data feeds that can support this type of metrology include the following: 
\begin{itemize}
    \item The Common Vulnerabilities and Exposures (CVE) list has 226,000 vulnerabilities (as of March 2024 and growing at approximately 25,000 per year) \cite{CVE}.
    \item The Common Weakness Enumeration (CWE) is a catalog of 938 weaknesses (as of March 2024) organized hierarchically \cite{CWE}. 
    \item The National Vulnerability Database (NVD) \cite{NVD} is a repository that provides a mapping of CVEs to CWEs and other services, such as providing attributes of vulnerabilities using the Common Vulnerability Scoring System (CVSS) \cite{CVSSv3} \cite{mell2022}.
    \item CWE View-1003 identifies a subset of the 130 ``most commonly seen weaknesses'' in CVEs \cite{View-1003}. For example, in 2023, View-1003 covered 94\,\% of the 85\,\% of published CVEs in NVD that had enough information to identify a CWE. NVD analysts use View-1003 for CWE analysis.
    \item  The Exploit Prediction Scoring System (EPSS) is a machine learning service that publishes the probability of a CVE vulnerability being exploited in the next 30 days \cite{EPSS, EPSSModel, EPSSPaper1, EPSSPaper2}. It uses data feeds provided by eight (as of March 2024) commercial security monitoring companies.
\end{itemize}

This paper provides an indirect statistical measurement of the exploitation of weaknesses in the wild. It introduces a simple metric to leverage a mashup of the aforementioned public resources to determine the probability of a CWE View-1003 weakness being actively exploited in a 30-day window. For a chosen CWE and date, the applicable CVEs are identified and their EPSS scores retrieved. A simple statistics equation is then used to determine the chance that at least one of these CVEs will be exploited. This yields the probability that the chosen CWE will be exploited in the wild within the subsequent 30-day window. 

For the empirical study, this measurement was conducted on all of the 130 View-1003 CWEs once a week between April 2021 and March 2024. The initial hypothesis was that all of the CWEs would be exploited in each 30-day window because of the large number of CVEs mapped to a small set of CWEs and the historically high levels of exploitation activity on the internet. 
An additional hypothesis was that deviations from this would randomly occur on a normal-like distribution. 

However, the analysis indicates that, for each 30-day window, only 8\,\% of the weaknesses are always being exploited. 43\,\% are exploited at least 90\,\% of the time (but not always). And 49\,\% are exploited less than 90\,\% of the time.
This means that for each such weakness and for significant periods of time, not a single vulnerability with the underlying weakness is projected to be observed to be exploited on the internet or enterprise networks.


Additionally, these CWEs usually have a probability history that consists of a sequence of identified common temporal patterns of exploitation. However, their probabilities do not act as a random variable with a normal distribution. Instead, they typically follow a series of patterns: drop, jump,
stable, and step up. 


The rest of this work is organized as follows. Section \ref{sec.metric} describes this work's metric for measuring the exploitation of weaknesses in the wild. Section \ref{sec.experiment} describes the experiment design in testing the metric. Section \ref{sec.results} presents the results and Section \ref{sec.discussion} discusses the results. Section \ref{sec.patterns} discusses the observed temporal patterns of weakness exploitation. Section \ref{sec.limitations} presents the limitations of the research.
Section \ref{sec.related} presents related work and Section \ref{sec.conclusion} concludes.

\section{Weakness Exploitation Metric}
\label{sec.metric}

Let the Probability Equation for CWE (PECWE) calculate the probability of a CWE being exploited in the wild within the next 30 days. More precisely, let PECWE be a function that takes a CWE $x$ and a date $d$ as input. It outputs the probability that some CVE with the underlying CWE $x$ will have been or will be observed to be exploited in the wild within the 30 days following date $d$.

For a CWE $x$, let $S_x$ be the set of all CVEs that are mapped to CWE $x$ by NVD.
For some CVE $y \in S_x$, let $EPSS(y,d)$ be the EPSS score for $y$ on date $d$. If $y$ is a CVE that has not been published by date $d$, then the value of $EPSS(y,d)$ is defined to be 0.

\[\text{PECWE}(x,d) = 1 - \prod_{\forall y \in S_x}^{}(1-EPSS(y,d)) \tag{1}\]
\label{Equation 1}

PECWE is a probability equation where the chance of success (i.e., CWE $x$ being observed to be exploited) is 1 minus the chance of all relevant events failing (i.e., all CVEs applicable to CWE $x$ not being exploited). PECWE works because EPSS scores are probabilities\cite{EPSSPaper1}. 

\section{Experiment Design}
\label{sec.experiment}

The experiment is designed to exercise the PECWE equation to determine its utility. Utility is evaluated by determining that PECWE values are not always pegged to 1.0 (the initial hypothesis) and that frequency (the number of CVEs applicable to a CWE) is not a simple replacement for PECWE.

The experiment is scoped by both a date range and the set of CWEs evaluated. The experiment date range is from April 14, 2021 (the EPSS initiation date), through  March 6, 2024 --- a total of 34.5 months. The set of CWEs evaluated was the 130 View-1003 CWEs (as of March 2023). NVD maps vulnerabilities to other CWEs but not in a systematic or comprehensive manner. It only includes such mappings when they are provided by external entities. Therefore, CWEs that are not in View-1003 are excluded. However, the special NVD CWE designators ``NVD-CWE-Other'' (``CWE not in View-1003'') and ``NVD-CWE-NoInfo'' (``insufficient information to make a determination'') are included, which makes for 132 data points. 

EPSS changed versions twice during the period of study. EPSS version 1 was released at the beginning of the study. Version 2 was released February 4, 2022. And version 3 was released March 7, 2023. Each version makes the same probability calculation but uses an improved machine learning model to improve accuracy. While each version represents an incremental improvement, the EPSS values of particular vulnerabilities can suddenly change between versions. This then results in some unusual PECWE probability changes on the version transition dates. This primarily affects the observed temporal patterns discussed in Section \ref{sec.patterns}. 

The set of CVEs applicable to each CWE was generated as follows. For each evaluated CWE, a traversal of the View-1003 hierarchy was performed to identify child CWEs. The NVD was then queried to retrieve the union of the applicable CVEs for the CWE and its children. This tree traversal and search at each child CWE was necessary because NVD maps each CVE to only the most specific CWE identifiable. In Equation 1, for a CWE $x$, $S_x$ is then a set formed from the union of the CVEs applicable to $x$ and all of its View-1003 children.

Note that it was not necessary to filter by the date when retrieving the CVEs applicable to each CWE. This is because a PECWE calculation for a CWE $x$ on a date $d$ may have CVEs in $S_x$ that do not yet exist, however, the EPSS score for those CVEs on date $d$ will be zero and have no effect on the PECWE output (each zero causes a PECWE intermediate result to be multiplied by one).

The PECWE probability for each CWE was calculated for the 132 data points on every Wednesday during the specified date range (since the first EPSS score was posted on a Wednesday). In the 34.5 months evaluated, there were 151 weeks. However, there are no data in EPSS for November 9, 2022. 
This set of data enabled the computation of 18,480 PECWE probabilities. 

\section{Results}
\label{sec.results}


This section provides results demonstrating that the PECWE equation meets the goal of the experiment. It shows that PECWE values are not always pegged to 1.0. It also shows that frequency, while correlated to PECWE, is not a replacement for it.

Subsection \ref{ssec.distributions} provides figures that show mean PECWE distributions.
Subsection \ref{ssec.range} describes the range of CWE PECWE probabilities over time. 
Subsection \ref{ssec.correlation} evaluates the correlation between the mean PECWE probabilities and the number of applicable CVEs.

Table \ref{tab:example_cwes} provides data for a set of example CWEs that illustrate the discussed statistics. The table contains a description of each CWE, the mean PECWE, the number of CVEs, and the PECWE range (presented in Section \ref{ssec.range}).

\input{DataPresentation/tab-examples}

\subsection{Distributions}
\label{ssec.distributions}

The mean PECWE for each CWE was calculated by taking the average of the weekly values. This enabled an abstracted single number summary for each CWE. Figure \ref{fig:distribution} shows the distribution of mean PECWE probabilities for the duration of the study for all 130 View-1003 CWEs and the two special CWE designators. Each x-axis value represents a CWE and is ordered by increasing mean PECWE probability. This shows that more than half of the CWEs do not have mean PECWE probabilities of 1.0.

Another important distribution is the number of published CVEs associated with each individual PECWE probability (not mean PECWE). Figures \ref{fig:scatter1} and \ref{fig:scatter2} show the number of CVEs associated with each weekly PECWE probability. Figure \ref{fig:scatter2} focuses on CWEs with 800 or fewer CVEs. Each data point represents a CWE on a specific date. The x-axis shows the number of CVEs utilizing that CWE, and the y-axis shows the PECWE. This shows that PECWE values can vary dramatically (e.g., from 0.1 to 1.0) for CWEs with an identical number of CVEs. This shows that frequency (number of CVEs) is not a replacement for PECWE.

\input{DataPresentation/fig-distributions}

\subsection{PECWE Probability Ranges}
\label{ssec.range}

The mean PECWE shown in Figure \ref{fig:distribution} is a useful abstraction, but it is also necessary to examine how the weekly PECWE values change over time. This data are shown in Table \ref{tab:range}. The 10 CWEs (8\,\%) labelled 'Exploited' had PECWE probabilities equal to 1.00 for the duration of the study. The 56 CWEs (43\,\%) labelled 'High' had PECWE probabilities greater than 0.90 but less than 1.00. The 63 CWEs (48\,\%) labelled 'Variable' had PECWE probabilities less than 0.90 and greater than to 0.10. And the 1 CWE (0\,\%) labelled 'Low' had PECWE probabilities less than or equal to 0.10. 


The variable CWEs often exhibit significant temporal patterns with the PECWE probabilities (explored in Section \ref{sec.patterns}).

\input{DataPresentation/tab-range}

\subsection{PECWE Correlation to the Number of Applicable CVEs}
\label{ssec.correlation} 

Figures \ref{fig:scatter1} and \ref{fig:scatter2} show a trend where a higher number of CVEs indicates a higher mean PECWE. This is further illustrated by the examples in Table \ref{tab:example_cwes}. CWE-79 (Cross-site Scripting) has the most CVEs at 25\,177 and is Exploited (mean PECWE is 1.00). CWE-920 has the fewest CVEs (just 3) and is Low with the lowest mean PECWE (0.02). However, this is not always the case. CWE-273 has a higher mean PECWE than CWE-354 even though the former has only 19 CVEs and the latter has 97. 

An examination of Equation 1 confirms and explains this trend of more CVEs promoting a higher PECWE probability. The more CVEs applicable to a CWE (i.e., the greater the size of $S_x$), the greater the opportunity to increase the PECWE since the $(1-EPSS(S_{x,n},d))$ terms are multiplied together prior to subtracting the product from 1.

This apparent correlation can be tested empirically. However, the common Pearson correlation cannot be used as the data are not linearly related (i.e., a change in one variable does not signify a proportional change in the other variable). Spearman's Rho correlation is more applicable but only provides a partial picture by measuring the degree to which a relationship is monotonically increasing while ignoring rate of change. Spearman's correlation for the relationship between the mean PECWE and the number of applicable CVEs per CWE is 0.90. This means that an increase in the number of CVEs applicable to a CWE is strongly correlated but not necessarily proportional to an increase in the mean PECWE probability. While this evaluation was done on the mean PECWE, the correlation was also tested and holds for individual PECWE values on specific dates.






\section{Discussion}
\label{sec.discussion}

This section discusses our analysis of the data presented in Section \ref{sec.results}. Subsection \ref{ssec.discussion.distributions} discusses the mean PECWE distributions. Subsection \ref{ssec.discussion.ranges} discusses the probability ranges. Subsection \ref{ssec.discussion.correlation} analyses the correlation results.

\subsection{Distribution Discussion}
\label{ssec.discussion.distributions}

Figure \ref{fig:distribution} demonstrates that many of the CWEs are not being exploited all of the time.

Figures \ref{fig:scatter1} and \ref{fig:scatter2} visually show how the mean PECWE probabilities appear to be non-linear and highly influenced by the number of CVEs applicable to each CWE. This relationship is further explored in Section \ref{ssec.discussion.correlation}.

These figures also demonstrate that the View-1003 CWEs have a greatly varying number of CVEs associated with them. This seems to indicate that certain CWEs are much more likely to be accidentally introduced into code than others. 

Despite the utility of PECWE, CVE frequency is still a useful metric for determining CWE importance. It measures how many vulnerabilities of a CWE type are introduced into code. PECWE is a complementary metric in calculating how often a particular CWE is being exploited in the wild. Note that CVE frequency could be used as an approximation of PECWE since the two are strongly correlated, but this is only accurate for CWEs with a high number of CVEs (since a high number of CVEs tends to peg the PECWE to 1.0).




\subsection{Probability Ranges Discussion}
\label{ssec.discussion.ranges}

120 CWEs (92\,\%) had PECWE probabilities less than 1.00 at some point during the study indicating that they are not always being exploited. 64 CWEs (49\,\%) had PECWE probabilities during the duration of the study that were less than 0.90. Figure \ref{fig:scatter2} shows that many of the individual PECWE values are low (indicating that active exploitation is unlikely for that 30-day window).  

\subsection{Correlation Discussion}
\label{ssec.discussion.correlation}

The 0.90 Spearman correlation between the mean PECWE probabilities and the frequency of applicable CVEs appears to indicate that the frequency of applicable CVEs could be used in place of PECWE probabilities. This is surprising because the vast majority of applicable CVEs do not or barely contribute to a CWE's PECWE calculation. 95\,\% of CVEs had an EPSS score less than 0.1 (using the EPSS dataset from October 4, 2023).

However, the Spearman correlation does not measure the proportionality of change, only that an increase in CVEs indicates an increase in PECWE probability. Figures \ref{fig:scatter1} and \ref{fig:scatter2} show that the PECWE relationship to the number of CVEs is extremely non-linear, rising almost vertically between 1 and 300 CVEs and then staying almost flat from 300 CVEs to 23,270. 
Thus, frequency is correlated to PECWE but is not a replacement for it in identifying which CWEs are actively exploited. They should be viewed as complementary. CVE frequency measures how often humans introduce bugs that cause particular CWEs. PECWE measures how often the introduced bugs for particular CWEs are exploited. While a high CVE frequency encourages a high PECWE, they are distinct. 

\section{Temporal PECWE Changes}
\label{sec.patterns}

This section examines how PECWE probabilities change over time. Section \ref{ssec.inter-EPSS-changes} examines the abrupt changes in PECWE probabilities that can occur when EPSS changes between versions. Section \ref{ssec.intra-EPSS-changes} examines the patterns of temporal changes that occur that are not related to EPSS version changes.

\subsection{Inter-EPSS Version Changes}
\label{ssec.inter-EPSS-changes}

EPSS changed versions and upgraded their machine learning model twice during the period of study: February 4, 2022, and March 7, 2023. The goal was to increase accuracy; this resulted in sudden, abrupt, and usually small changes to EPSS values on the version change dates. However, even a small EPSS probability change to a few vulnerabilities can result in a large change in PECWE probability. 

Consider the example CWEs from Table \ref{tab:example_cwes}. You can see their PECWE value changes between EPSS version numbers in Figures \ref{fig:CWE-79} through \ref{fig:CWE-920} (the black circles indicate the days on which version changes occurred). Nothing happened in Figure \ref{fig:CWE-79} during the version changes because the PECWE was strongly pegged to 1.0. And not all abrupt changes were caused by version changes.

\subsection{Intra-EPSS Version Patterns}
\label{ssec.intra-EPSS-changes}

Within the timeframe of an EPSS version, changes in PECWE are observed for the 130 view-1003 CWEs. They appear to follow a few different patterns as opposed to simply varying randomly. Each PECWE value series for a CWE appears to tell a story that is doubtlessly related to real-world events regarding vulnerability discovery, exploitation, and mitigation. 

An arbitrary threshold of 0.1 was used as the minimum probability change from which to identify a pattern. Many CWEs' weekly PECWE probabilities exhibit a combination of two or more patterns, and a particular pattern may apply to only part of a time series. Many CWE time series can be described as a sequence of these patterns. The four common patterns of CWE exploitation identified are as follows: 

\begin{enumerate}
    \item Drop: The probabilities precipitously fall.
    \item Jump: The probabilities precipitously rise.
    \item Stable: The probabilities stay almost the same.
    \item Step Up: The probabilities regularly and gently ascend, mostly monotonically.
\end{enumerate}

The pattern of Step Down (i.e., a monotonic gentle decrease, opposite of Step Up) was analyzed, and --- non-intuitively --- none of the 130 CWEs exhibited this pattern. This was not only true for the entire time series for each EPSS version, CWE but also for smaller periods. 



As discussed in Section \ref{ssec.range} and shown in Table \ref{tab:range}, there are 66 CWEs that have a weekly PECWE probability range descriptor of either High or Exploited and one that is Low. They all have the overall pattern of Stable. However, many contain minimal (less than a 0.1 change) Step Ups, Drops, and Jumps.

For example, Figure \ref{fig:CWE-79} shows the pattern Stable for the Exploited CWE-79 (Improper Neutralization of Input During Web Page Generation, or ``cross-site scripting''). 
Also, Figure \ref{fig:CWE-920} shows the pattern Stable for the Low CWE-920 (Improper Restriction of Power Consumption).

The patterns of the 63 CWEs with a Variable range descriptor (see Table \ref{tab:range}) are of more interest. These all have patterns that exhibit significant probability changes. 

Figure \ref{fig:CWE-273} shows that Variable CWE-273 (Improper Check for Dropped Privileges) has a Jump at the end of the EPSS version 1 time period. It has a Step Up in the middle of the EPSS version 3 time period.

Figure \ref{fig:CWE-354} shows that Variable CWE-354 (Improper Validation of Integrity Check Value) has a Drop followed by a Step Up during the EPSS version 2 time period. It has a large Jump at the end of the EPSS version 3 time period.

Figure \ref{fig:CWE-367} shows that Variable CWE-367 (Time-of-check Time-of-use (TOCTOU) Race Condition) has a Step Up during the EPSS version 2 time period.

Figure \ref{fig:CWE-697} shows that Variable CWE-697 (Incorrect Comparison) has a Jump at the end of the EPSS version 1 time period. It has a Step Up in the middle of the EPSS version 2 time period.

\input{DataPresentation/fig-patterns}
\section{Limitations}
\label{sec.limitations}

EPSS primarily uses network intrusion detection sensors (IDSs) to create its training data. This could limit its ability to predict CVE exploitation for CVEs that are only detected through other detection mechanisms (e.g., host-based IDSs). Such CVEs may have similar CWEs, resulting in the inability of PECWE to predict exploitation those CWEs. While not enough data are available to analyze this potential limitation, none of the PECWE time series for the CWEs pegged at 0\,\%. This indicates that the EPSS sensors are detecting at least some CVEs for all evaluated CWEs.  



\section{Related Work}
\label{sec.related}

PECWE is a method to identify important CWEs that should be targeted by the security community for mitigation and elimination. Another approach is to identify ``dangerous'' CWEs using the MITRE CWE Top 25 Most Dangerous Software Weaknesses \cite{CWETOP25}. This is an annually updated public list of CWEs that are touted to have associated vulnerabilities that are both large in number (frequency) and severe (using CVSS). In \cite{galhardo2020}, it was discovered that the Top 25 equations are biased toward frequency, and the included severity component is minimized. Even though frequency is normalized in the Top 25, the maximum to minimum range is so large that almost all CWEs are on the extreme low end of the frequency range. Thus, the Top 25 list is largely a measure of the number of CVEs associated with each CWE (i.e., frequency) \cite{b2}.

CVE frequency is a useful metric for measuring the importance of CWEs because it shows which CWEs developers are most introducing into code. PECWE should not replace frequency. However, since only ``2\,\% of published vulnerabilities have observed exploits in the wild'' \cite{CyentiaReport}, the vast majority of introduced vulnerabilities appear to have no impact on security. This statistic argues for an approach to identify dangerous CWEs by their use in the wild. As mentioned in the introduction, the data to directly measure this are not available. However, PECWE is an indirect measurement that can be used until more direct measurements are available.

\section{Conclusion}
\label{sec.conclusion}




With over 227\,000 known vulnerabilities in software, an additional 25\,000 being discovered every year, it is non-intuitive that only 8\,\% of the weaknesses are observed to be exploited in every 30-day window. We also find it unexpected that 49\,\% are exploited less than 90\,\% of the time. This is especially surprising given the high levels of exploitation regularly conducted via the internet by independent hackers, corporate espionage agents, organized crime, and nation-states. 

The security community has identified only 130 weaknesses that account for the vast majority of discovered vulnerabilities \cite{View-1003} --- more than three orders of magnitude smaller than the number of known vulnerabilities. This means that most vulnerabilities are due to 130 types of bugs or faults that software developers are introducing into their code and that manual code reviews and security code scanners are not fully catching. While still a non-trivial amount, this number reduces the seeming hopelessness of preventing and detecting the 25,000 vulnerabilities published every year.

This work further reduces the complexity of addressing weaknesses in code by demonstrating that the majority of weaknesses are not constantly exploited. For significant periods of time, the probabilities indicate that not a single vulnerability with such underlying weaknesses was observed to be exploited on the internet or enterprise networks. This implies that when such a weakness is being exploited, the number of associated vulnerabilities and the severity of exploitation may be small.

The development of the PECWE metric assists in prioritizing efforts to eliminate vulnerabilities of particular weakness types. It differentiates and highlights the weaknesses that are almost always being exploited from those that are not.

%% file: DataPresentation/tab-examples.tex
\begin{table*}
\caption{Example Set of Representative CWEs}
\label{tab:example_cwes}
\begin{tabular}{lllll}
CWE & 
Name &
\begin{tabular}[c]{@{}l@{}}Mean\\ PECWE\end{tabular} & 
\begin{tabular}[c]{@{}l@{}}Number\\ of CVEs \end{tabular} &
\begin{tabular}[c]{@{}l@{}}PECWE\\Range \end{tabular} \\
\hline
CWE-79  & \begin{tabular}[c]{@{}l@{}}Improper Neutralization of Input During Web Page \\ Generation ('Cross-
site Scripting')\end{tabular} & 1.00 & 25\,177 & Exploited  \\
CWE-273 & Improper Check for Dropped Privileges & 0.51 & 19 & Variable \\
CWE-354 & \begin{tabular}[c]{@{}l@{}}Improper Validation of Integrity Check Value\end{tabular} & 0.41 & 97 & Variable\\
CWE-367 & \begin{tabular}[c]{@{}l@{}}Time-of-check Time-of-use (TOCTOU) Race Condition\end{tabular} & 0.7 & 274 & Variable \\
CWE-697 & Incorrect Comparison & 0.73 & 103 & Variable \\
CWE-920 & Improper Restriction of Power Consumption & 0.02 & 3 & Low \\
\end{tabular}
\end{table*}

%% file: DataPresentation/fig-distributions.tex
\begin{figure}
\centerline{\includegraphics[scale=.65]{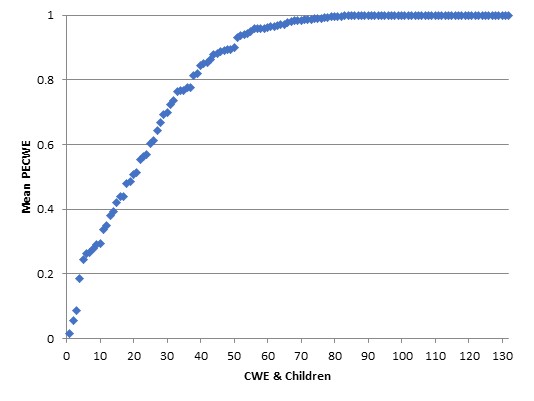}}
\caption{Distribution of mean PECWE values for the View-1003 CWEs from April 14, 2021, to March 6, 2024}
\label{fig:distribution}
\end{figure}


\begin{figure}
\centerline{\includegraphics[scale=.65]{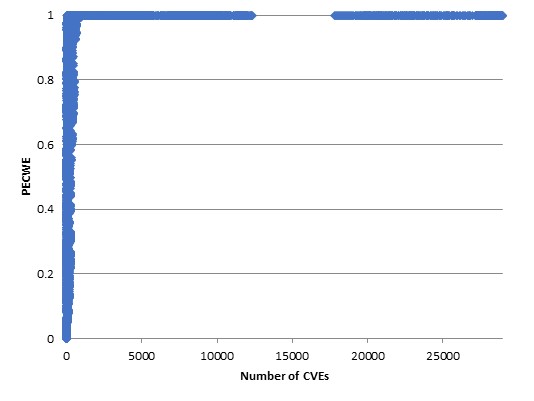}}
\caption{Number of CVEs Associated with each PECWE Score for View-1003 CWEs 
}
\label{fig:scatter1}
\end{figure}

\begin{figure}
\centerline{\includegraphics[scale=.65]{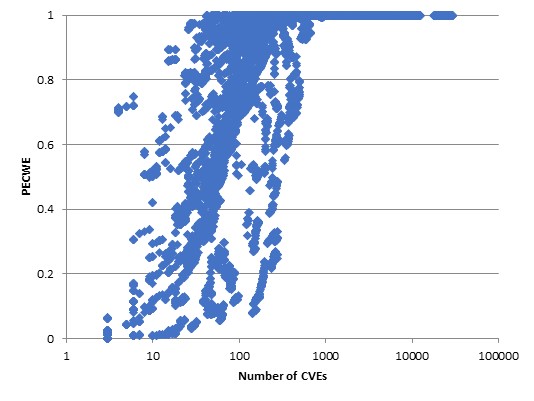}}
\caption{Number of CVEs Associated with each PECWE for View-1003 CWEs -- with log$_{10}$ x-Axis 
}
\label{fig:scatter2}
\end{figure}

%% file: DataPresentation/tab-range.tex
\begin{table}
\caption{Range of CWE PECWE Probabilities \\ During the Period of Study 
}
\label{tab:range}
\begin{tabular}{llll}
\begin{tabular}[c]{@{}l@{}}Range\\Descriptor\end{tabular} & Range & Count & List of CWEs \\

\hline
Exploited & =1.00 & 10 & 
\begin{tabular}[c]{@{}l@{}}
20 22 78 79 119 125 200 416 \\
502 787
\end{tabular}\\

\hline
High & \textgreater{}=0.90 & 56 & 
\begin{tabular}[c]{@{}l@{}}
59 74 77 88 89 91 94 120 134 \\
190 203 209 269 276 287 295 \\
306 319 326 330 352 362 369\\
400 401 404 415 426 427 434\\
444 476 522 532 552 601 611\\
617 665 674 704 732 755 770\\
772 798 824 835 843 862 863\\
908 917 918 1321 \\
\end{tabular} \\

\hline
Low & $<$=0.10 & 1 & 920\\

\hline
Variable & \begin{tabular}[c]{@{}l@{}}Not in\\ another\\ category\end{tabular} & 
63 &
\begin{tabular}[c]{@{}l@{}}
116 129 131 178 191 193 212 \\
252 273 281 290 294 307 311 \\
312 327 331 335 338 345 346 \\
347 354 367 384 407 425 428 \\
436 459 470 494 521 565 610 \\
613 639 640 662 667 668 669 \\
670 672 681 682 697 706 754 \\
763 776 829 834 838 909 913 \\
916 922 924 1021 1236 1284 \\
1188 1333
\end{tabular}\\

\end{tabular}
\end{table}


%% file: DataPresentation/fig-patterns.tex

\begin{figure}
\centerline{\includegraphics[scale=.6]{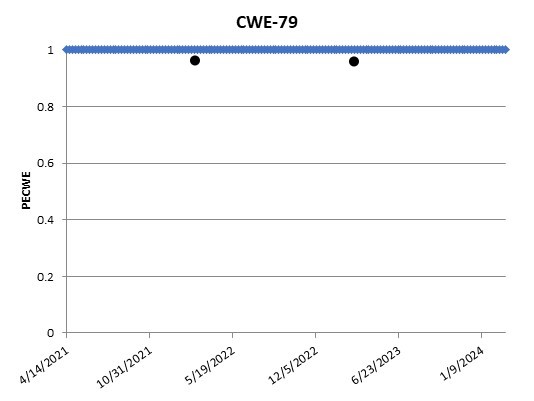}}
\caption{CWE-79 PECWE Probabilities\\ (EPSS version number changes marked by black dots)}
\label{fig:CWE-79}
\end{figure}

\begin{figure}
\centerline{\includegraphics[scale=.6]{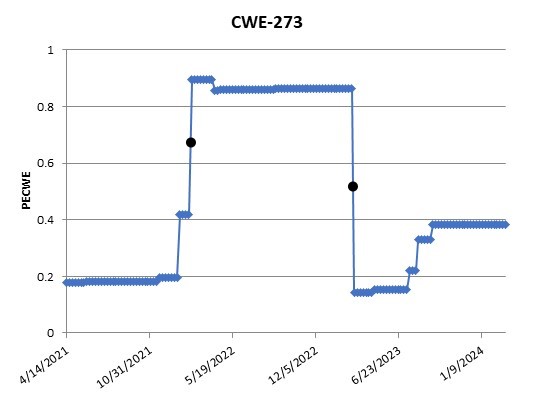}}
\caption{CWE-273 PECWE Probabilities\\ (EPSS version number changes marked by black dots)}
\label{fig:CWE-273}
\end{figure}

\begin{figure}
\centerline{\includegraphics[scale=.6]{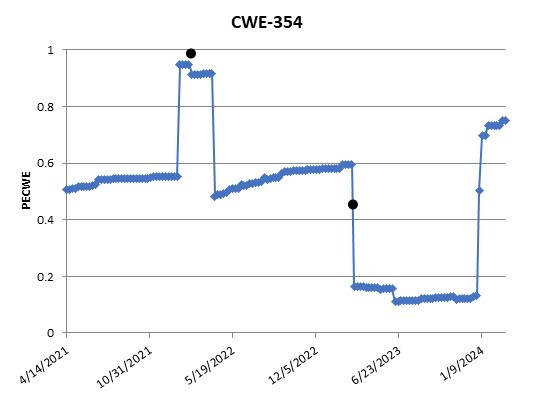}}
\caption{CWE-354 PECWE Probabilities\\ (EPSS version number changes marked by black dots)}
\label{fig:CWE-354}
\end{figure}

\begin{figure}
\centerline{\includegraphics[scale=.6]{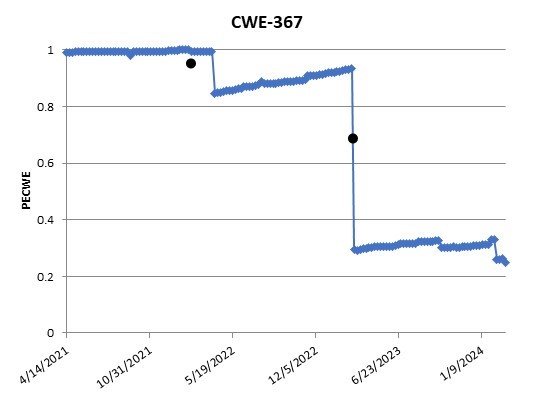}}
\caption{CWE-367 PECWE Probabilities\\ (EPSS version number changes marked by black dots)}
\label{fig:CWE-367}
\end{figure}

\begin{figure}
\centerline{\includegraphics[scale=.6]{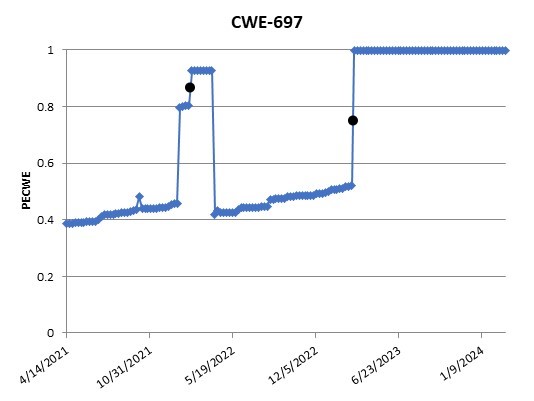}}
\caption{CWE-697 PECWE Probabilities\\ (EPSS version number changes marked by black dots)}
\label{fig:CWE-697}
\end{figure}

\begin{figure}
\centerline{\includegraphics[scale=.6]{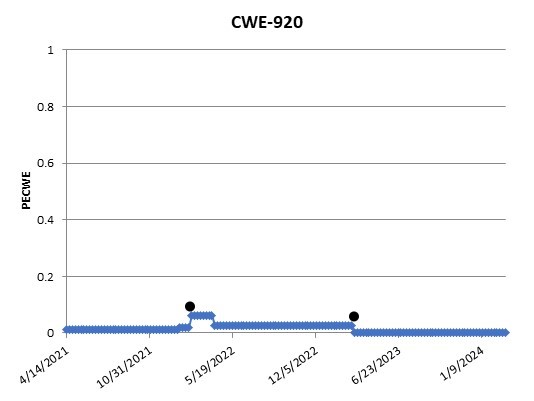}}
\caption{CWE-920 PECWE Probabilities\\ (EPSS version number changes marked by black dots)}
\label{fig:CWE-920}
\end{figure}

%% file: references.tex
\vspace{12pt}